\documentclass{article}

\usepackage{microtype}
\usepackage{graphicx}
\usepackage{subcaption}
\usepackage{booktabs} 
\usepackage{array}

\usepackage{xurl}
\usepackage{hyperref}

\usepackage[accepted]{icml2026}

\usepackage{amsmath}
\usepackage{amssymb}
\usepackage{mathtools}
\usepackage{amsthm}

\usepackage[capitalize,noabbrev]{cleveref}

\theoremstyle{plain}

\theoremstyle{definition}

\theoremstyle{remark}

\usepackage[textsize=tiny]{todonotes}

\icmltitlerunning{Verifying Restrictions on Frontier AI Research}

\begin{document}

\twocolumn[
  \icmltitle{Verifying Restrictions on Frontier AI Research}




  \begin{icmlauthorlist}
    \icmlauthor{Aaron Scher}{yyy}
  \end{icmlauthorlist}

  \icmlaffiliation{yyy}{Technical Governance Team, Machine Intelligence Research Institute, Berkeley, USA}

  \icmlcorrespondingauthor{Aaron Scher}{aaron.scher@intelligence.org}

  \icmlkeywords{AI Verification, AI Governance, AI Agreements}

  \vskip 0.3in
]



\printAffiliationsAndNotice{}  

\begin{abstract}
The premature development of artificial superintelligence poses major risks to humanity, so researchers have proposed international agreements halting such development until it can be done safely. AI progress depends primarily on compute, algorithms, and data; a durable halt would address all three so that advances in one input do not counteract restrictions on another. Improvements to AI algorithms are driven largely through research activities, so this research may need to be restricted during a halt. Given low international trust, signatories will want to verify compliance. This paper analyzes how such restrictions on AI research could be verified, while remaining agnostic about what specific research would be prohibited. It first explores key considerations that affect the verifiability of research restrictions, such as the computational infrastructure necessary for experiments. It then catalogs 28 candidate verification mechanisms. These mechanisms include whistleblowers, search warrants, reviews of AI training code, standard intelligence gathering tools, and more. Some of these mechanisms are not yet implementation-ready, and some might be undesirable upon further inspection. By examining the space of potential options, this work provides a foundation for future research to develop the most promising mechanisms into deployable tools.
\end{abstract}

\section{Introduction}

Many experts in the field of artificial intelligence (AI) worry that increasingly intelligent AI systems will pose catastrophic risks to humanity, including human extinction \citep{center_for_ai_safety_statement_2023,yudkowsky_if_2025}. To address this threat, some have proposed delaying the development of artificial superintelligence (ASI) until development can be done safely \citep{future_of_life_institute_statement_2025,barnett_ai_2025,scher_international_2025,ramiah_toward_2025,martin_analysis_2024}.

To this end, \citet{scher_international_2025} provide a draft international agreement aimed at preventing the premature creation of ASI\@. For simplicity, we use the agreement from \citet{scher_international_2025} as a scaffold for this work, asking how research restrictions could be verified under that agreement; however, these mechanisms would also be useful in other policy proposals that involve research restrictions. The agreement has two main pillars focused on stopping the development of better AIs: \textit{chip controls} and \textit{research restrictions}. Both of these are needed; neither would be sufficient on their own. The primary drivers of AI progress are compute, data, and algorithms \citep{buchanan_ai_2020}. If only compute is controlled, progress would still continue via advances in data and algorithms \citep{ho_algorithmic_2024,hooker_limitations_2024,heim_training_2024}. The proposed agreement includes research restrictions to fill this gap, focused primarily on research that is likely to advance the state of frontier AI capabilities (e.g., frontier AI research, general-purpose AI capabilities research, potentially AI chip design research). Certain research would also endanger the verification system established in the agreement, so it may be desirable to prohibit such research as well (e.g., decentralized training methods). The question of which specific research should be restricted during a halt is out of scope for this paper.

We focus instead on the verification question: \textbf{How could governments verify compliance with restrictions on AI research, especially within signatory countries?} We answer this question in two ways:

\begin{itemize}
    \item \textbf{First, we analyze key factors influencing the verifiability of research restrictions.} These include the need for compute in testing research ideas, the present rarity of human experts who can push the frontier, and the likely automation of AI research activities.
    \item \textbf{Second, we map the space of potential verification mechanisms, cataloging 28 high-level mechanisms} that could be used to verify compliance with research restrictions. Examples of these mechanisms include whistleblowers, automated code reviews, and sting operations.
\end{itemize}

To simplify, we use specific terminology throughout this paper. \textit{Restricted research} refers to research that is prohibited by an international agreement, for instance because it contributes to AI capabilities or threatens the verification system; this paper remains agnostic about where the boundaries of that category should be drawn. \citet{scher_international_2025} use restricted research as an umbrella term covering two forms: \textit{controlled} research, which is subject to monitoring, and \textit{banned} research, which is disallowed. In this paper, we use restricted research to refer to the latter. \textit{High-risk organizations and researchers} are organizations and individuals that are relatively likely to carry out impactful restricted research, for instance because their legitimate work is similar to restricted research (i.e., it falls in the controlled category) or because they have relevant expertise and resources. For example, current AI developers such as OpenAI and Anthropic would qualify as high-risk organizations due to their expertise and resources. After an agreement, they would not be allowed to continue conducting frontier capabilities research, but they could, for instance, serve existing models.

The \textit{verification authorities} are a group of people responsible for verification, for instance a technical body staffed by representatives from certain signatory countries, as in the agreement from \citet{scher_international_2025} (though we aim to be agnostic to the design of such a body as this is primarily a political rather than technical question). The choice of verification authorities affects the political viability and security risk of mechanisms, but these considerations will be balanced against countries' need to obtain justified confidence in compliance, not trusting each other to self-enforce an agreement. \textit{Inspectors} are employees of the verification authority who carry out specific monitoring or inspection tasks. Finally, we use the term \textit{verification} to refer to a broad set of methods and actions that could make one party justifiably more confident that another is adhering to agreed-upon rules. This is somewhat broader than the standard use as it includes preventative mechanisms and useful components as well as core information collection---we note when our use is more broad than typical.

To ground this work, our investigation is motivated by three main threat actors who may attempt to violate research restrictions. First, covert nation-state actors who are highly competent at security, may have secret data centers, and whose existence may not be known to verification authorities (e.g., a secret Manhattan Project for ASI). Second, established organizations that are known to verification authorities, are highly competent at AI development, and are generally law-abiding (e.g., OpenAI, DeepSeek, Prime Intellect). Third, rogue academics and less-resourced researchers who may be ideologically motivated and are likely to make progress either slowly (and with contributions from many individuals) or through new paradigm advances, if at all. In some cases, the potential violator is known, so verification would involve countries agreeing to monitoring and inspections to verify compliance. In other cases the main verification difficulty is to detect the presence of such a project.

\section{Key considerations for the verifiability of research restrictions}

\textbf{Compute is needed to test many research ideas} within the machine learning paradigm. \citet{barnett_compute_2025} investigates how much compute was used to develop the algorithmic innovations used in Llama 3 and DeepSeek-V3 and finds that around half of the innovations were developed using less than 8 H100 equivalents of computing capacity (or around $1\times 10^{21}$ FLOP when measuring total computations). This quantity is larger than what average consumers own, but small enough to be effectively undetectable: we are not aware of any detection mechanism that could reliably find all such clusters in the world, though such an analysis is outside the scope of this work.

\textbf{Validating research ideas at scale requires substantial compute.} Industry practitioners report that many ideas which perform well in small experiments nonetheless scale poorly, and to identify promising directions requires larger-scale experiments \citep[00:45:43]{dean_jeff_2025}. Recent work by \citet{gundlach_origin_2025} finds that several of the most consequential algorithmic innovations are compute-dependent: their benefit differs when applied at small vs.\ large training scales. Algorithm researchers seeking to validate such ideas and push the capability frontier would therefore require more compute, rendering their activity more detectable.

The extent to which frontier AI research is bottlenecked by compute is not publicly established. Prior work examining the implications of automated AI research has diverged on the severity of this bottleneck \citep[p.~44]{anthropic_system_2026}; see also \citet{whitfill_will_2025} and \citet{davidson_will_2025}.

\textbf{The population of researchers relevant to frontier AI development is small—plausibly on the order of thousands or tens of thousands}. \citet{scher_international_2025} provide rough estimates for the number of these researchers, ranging from thousands (i.e., technical staff at frontier AI companies) to tens of millions (i.e., all software engineers and AI chip company employees), with tens of thousands being a reasonable overall estimate. The skills required to advance the frontier are rare. In a world where humans, rather than AI systems, remain the primary drivers of AI research progress, the set of individuals whose activities warrant monitoring is correspondingly bounded.

\textbf{AI systems may come to perform most AI research in the future.} At the time of writing, frontier AI companies are working to automate the labor of their human engineers and researchers \citep{ecoffet_event_2026,reedi_ml-intern_2026}, though humans remain important in the process. This balance is likely to shift substantially within the next several years \citep{favaro_when_2026,altman_built_2026}. If and when an international agreement enters into force, AI systems will likely conduct a substantial share of AI research and development---perhaps all of it. This shift cuts in both directions for verification. On one hand, AI systems are trivially copyable and considerably harder to detect than human researchers, especially smaller models capable of running on consumer hardware. On the other hand, imposing intensive monitoring regimes on AI systems raises fewer normative and legal objections \citep{mark_constitutional_2025}.

\textbf{Criminalization would substantially reduce willingness to engage in affected research.} Under the agreement proposed by \citet{scher_international_2025}, specified categories of AI research would become illegal and would likely acquire significant social stigma. The agreement does not specify enforcement mechanisms in detail, requiring only that each party ``prohibits and prevents'' restricted research; in practice, penalties could plausibly include prison sentences. Individuals and corporations typically comply with legal prohibitions for a range of reasons, including the expected cost of detection and sanction \citep{nagin_deterrence_2013}. Investment in the affected areas would likewise contract in response to heightened legal and financial exposure. As an informal estimate, we expect aggregate effort directed toward frontier AI research to fall to approximately 1--10\% of its pre-agreement level, though this figure is speculative and sensitive to the credibility of enforcement.

\section{Mechanisms that could be used to verify compliance with research restrictions}

We catalog 28 of the most promising high-level mechanisms from among a larger set considered. The full set draws on precedent, related fields, and an understanding of the AI R\&D process. We group mechanisms by their target---identifying covert projects, monitoring declared high-risk organizations, verifying use of declared chips, or broad non-proliferation. We then sort each group in descending order of priority---an informal assessment of the costs (e.g., privacy concerns, amount of R\&D needed, difficulty of implementation) and benefits (e.g., effectiveness against the threat actors). The least promising approaches are omitted entirely.

\begin{table}[!htbp]
\centering
\caption{Catalog of 28 candidate verification mechanisms for restrictions on AI research, grouped by their target.}
\label{tab:mechanisms}
\small
\renewcommand{\arraystretch}{1.2}
\newcolumntype{L}{>{\hangindent=1.5em\hangafter=1}p{0.92\columnwidth}}
\begin{tabular}{L}
\toprule
\textit{Covert projects} \\
\midrule
Whistleblowers \\
Intelligence gathering on known high-risk targets \\
Intelligence gathering to identify unknown high-risk targets \\
Interviews with researchers and relevant officials \\
International search warrants and inspections \\
Leadership emphasis on compliance$\dagger$ \\
Polygraphs and lie detection \\
Sting operations for covert projects \\
\midrule
\textit{Declared high-risk organizations} \\
\midrule
AI-assisted code review for compliance \\
Auditors in high-risk organizations \\
Employee-monitoring software overseen by inspectors \\
Monitoring the use of chip design tools \\
Monitoring chip fabrication \\
Multi-party compliance attestations \\
Pre-registration and review of controlled research plans \\
Pre-publication review of controlled research \\
Researcher education and outreach$\dagger$ \\
Recordkeeping and disclosure on request \\
Evaluating AI models to detect unexplained capability gains \\
Controlled work environments to prevent evasion \\
\midrule
\textit{Chip-use verification for declared chips} \\
\midrule
Review AI training for novel methods of relevance \\
Monitoring AI inference content for restricted research \\
Review AI training data for prohibited data improvements \\
Broad chip-use monitoring for restricted experiments \\
Restricting chips to only run approved models \\
\midrule
\textit{Non-proliferation} \\
\midrule
Preventing the development of AI-R\&D-capable AIs$\dagger$ \\
Preventing the proliferation of AI-R\&D-capable AIs$\dagger$ \\
Non-proliferation of AI algorithms and data$\dagger$ \\
\bottomrule
\end{tabular}

\vspace{0.3em}
\begin{minipage}{0.92\columnwidth}
\footnotesize
$\dagger$ Not a verification mechanism in the traditional use of the term, but still useful for obtaining confidence in compliance.
\end{minipage}
\end{table}

\subsection{Covert projects}

During an international agreement on AI, actors may seek to continue prohibited development in secret. The mechanisms below can help detect, identify, or prevent such covert projects.

\textbf{Whistleblowers} Establish secure channels for whistleblowing about the existence of secret AI projects, including anonymous channels. If possible, establish strong incentives around whistleblowing. The SEC's Dodd-Frank whistleblower program, which offers 10--30\% of collected sanctions and has paid over \$2 billion since 2010, provides a direct model for financially incentivized reporting with anonymity and anti-retaliation protections. Asylum guarantees that extend to a whistleblower’s family may also incentivize reporting.

\textbf{Intelligence gathering on known high-risk targets} Use traditional intelligence gathering methods to monitor high-risk organizations and individuals. States might employ such methods as signals and communications intelligence to understand whether restricted research projects are taking place. They might also use financial intelligence to detect AI chip rentals or hiring, and human intelligence, including interviews with researchers and those close to them.

\textbf{Intelligence gathering to identify unknown high-risk targets} Intelligence agencies search for covert projects by first identifying individuals and organizations that are likely to engage in restricted research. The methods used might parallel those used in the past to detect covert WMD projects, including human intelligence and signals intelligence \citep{committee_of_privy_counsellors_review_2004}. Compared to historical analogues, AI research appears more difficult to detect: data centers and chips are the only input with a notable physical footprint, and AI research requires relatively few chips. Arms control treaties from SALT I to New START explicitly authorize national technical means for verification and prohibit parties from interfering with each other's monitoring capabilities. In some cases, actors who continue AI research despite global restrictions might leave public clues. They might publish their research findings online (e.g., for ideological reasons), organize events, or otherwise signal non-compliance. Therefore, open source intelligence and monitoring public AI research might be effective against non-covert rulebreakers. The IAEA systematically collects and analyzes open source scientific literature as part of its safeguards verification activities, and this practice can cleanly adapt to monitoring public AI research.

\textbf{Interviews with researchers and relevant officials} Verification authorities interview high-risk researchers, aiming to uncover whether the researchers performed restricted research after the restrictions took effect. To guard against misconduct, a researcher’s country of origin might oversee the interview. Interviews are a precedented measure; the Chemical Weapons Convention (CWC) Verification Annex explicitly grants inspectors the right to interview facility personnel.

\textbf{International search warrants and inspections} Carry out search warrants for properties, files, and computers, a standard method for gathering evidence about suspected illegal activity. Information collected from these sources can identify whether the targets are engaged in restricted research. Research is difficult, and researchers benefit from recording their ideas and experiments, often creating a paper trail. Warrants and inspection bodies would likely need to be international to avoid lax self-policing or overreach by foreign authorities. Possible design features include approval from an international judge appointed with the consent of the host country. To guard against collusion with or tipoffs from domestic sources, warrants could follow the precedent set by the CWC, which established a right to short-notice challenge inspections at any facility in any state party, with no right of refusal, although this right has not yet been exercised.

\textbf{Leadership emphasis on compliance} Major world leaders publicly declare that participation in restricted research or covert frontier AI development is a serious offense, even if directed by a superior. Leaders encourage whistleblowing, discourage restricted AI research, and emphasize the seriousness of this position at all levels of command. CIA Director Schlesinger's 1973 directive \citep{schlesinger_memorandum_1973} ordering all employees to report activities outside the Agency's charter is a precedent for senior leaders using explicit directives to signal that compliance is a top priority. This action could make covert projects less likely and would be visible to rivals, but it is not a verification measure in the traditional use of the term.

\textbf{Polygraphs and lie detection} Sponsor research into lie detection technology and apply it as an added layer of screening in conversations with high-risk researchers and with public officials likely to know about covert projects. Improvements in lie detection would render interviews, meetings, and public statements more credible. Polygraph examinations are required for all CIA and NSA personnel, but the scientific community has not established that they are reliable screening tools \citep{national_research_council_polygraph_2003}. Even weak or unreliable signals could be useful to prompt further investigation.

\textbf{Sting operations for covert projects} Run sting operations or false-flag recruitment to catch researchers willing to unlawfully participate in a covert project, chip owners willing to rent or sell to such projects, or other nascent threats. Sting operations have been used by law enforcement in domains ranging from drug enforcement to terrorism and WMD procurement.

\subsection{Declared high-risk organizations}

To aid verification, an agreement would likely require that organizations and individuals that pose a significant risk of conducting restricted research declare themselves to the verification authorities and accept monitoring protocols such as those below.

\textbf{AI-assisted code review for compliance} Require that all production and research code at high-risk organizations be screened by an AI to flag possible restricted research. Automated screening might extend to codebases, experiment logs, and research documents, and might output only compliance-relevant findings (e.g., ``this code implements a novel training algorithm, thus constituting restricted research'') without revealing proprietary details to human inspectors. One difficulty is in third-party inspectors verifying that the right code review happens every time it needs to, and another is ensuring the AIs doing code review will do a good job (i.e., they are not jailbreakable, are not attempting to sabotage verification, and are good at the relevant skills). Automated tools such as linters and SAST are common in coding, and AI tools such as Claude Code Review and Cursor Bugbot also see growing adoption. While these tools traditionally focus on bugs and security vulnerabilities, some of them could be adapted to detect violations of an agreement.

\textbf{Auditors in high-risk organizations} Maintain full-time human auditors, provided by the verification authorities, within select high-risk organizations. Auditors aim to ensure that the organization is not engaging in restricted research. The Nuclear Regulatory Commission has stationed full-time resident inspectors at every U.S. nuclear power plant since 1978 to observe daily operations and verify compliance \citep{us_nuclear_regulatory_commission_resident_2025}.

\textbf{Employee-monitoring software overseen by inspectors} Require employee-monitoring software on the work devices of high-risk organizations and researchers. Potential implementations vary in the data they collect and how it is analyzed. Inspectors would need sufficient access to verify compliance, but could analyze data using automated methods to preserve privacy. Similar software is widespread in corporate and classified government settings, including data loss prevention software, endpoint detection and response, and application-specific user activity logging, though having external inspectors oversee such monitoring on researchers' devices is a novel addition to existing practice.

\textbf{Monitoring the use of chip design tools} Collaborate with Electronic Design Automation (EDA) companies (Synopsys, Cadence, Siemens EDA) to ensure their tools are not being used in ways that violate the agreement. Semiconductor foundries (e.g., TSMC, Intel) provide designers with a Process Design Kit (PDK) specific to their fab and necessary for fabrication. These companies could work with verification authorities to limit tooling use to legitimate organizations performing only permitted research. This mechanism is relevant if research restrictions include research into better AI chips.

\textbf{Monitoring chip fabrication} Collaborate with chip fabrication companies to ensure new chips are not the product of restricted research. Cutting-edge chips are fabricated by a small handful of companies (e.g., TSMC, Intel, SMIC, Samsung) who have significant visibility into chip design. Various performance characteristics of chips could be inferred from either chip designs or fabricated chips (e.g., via independent teardown analysis), so the world's chip production could be monitored to ensure that chips are not advancing along the relevant axes---for agreements in which research on improving frontier AI chips is restricted.

\textbf{Multi-party compliance attestations} Require high-risk organizations to periodically submit attestations from multiple independent employees (e.g., 5+ signatories from different divisions) asserting that no restricted research is active or planned. This is cheap to implement but would be ineffective against a highly competent adversary. The Sarbanes-Oxley Act \citep[Sections~302 and~906]{united_states_congress_sarbanes-oxley_2002} provides a model for executive-level compliance attestations, requiring CEOs and CFOs to personally certify financial reports each quarter and specifying criminal penalties for false statements.

\textbf{Pre-registration and review of controlled research plans} Require high-risk organizations and researchers to pre-register their research plans with verification authorities. Pre-registration provides early screening of research proposals likely to trigger restrictions, and could be automated to minimize added friction. This approach primarily prevents restricted research from happening by accident, and may also reduce plausible deniability for bad actors: violations either result from a classification error by the verification authorities, or from bad actors circumventing the registration process, itself a flagrant violation. One difficulty in implementing this mechanism is that predicting downstream effects of research is inherently difficult. Pre-registration and review of research exists in the U.S. for human subjects research (Institutional Review Boards) and clinical trials (U.S. Food and Drug Administration), though these processes are much slower than what could be achieved with automation.

\textbf{Pre-publication review of controlled research} Require high-risk organizations and researchers to submit their research results to verification authorities before publication. Verification authorities would check that the work does not constitute restricted research, potentially using automated classifiers to speed up the process. The Atomic Energy Act classifies nuclear weapons design information by default \citep{united_states_congress_atomic_1954}, and the Invention Secrecy Act of 1951 authorizes the government to prohibit publication of patent applications with national security implications \citep{united_states_congress_invention_1952}. In contrast, computer science research is sometimes conducted semi-publicly, for instance using public code repositories, and for work in high-risk categories, this practice would need to change.

\textbf{Researcher education and outreach} Interface with high-risk researchers to ensure they understand the research restriction rules, the consequences of non-compliance, and available whistleblower pathways. This would be relatively cheap to implement and would strengthen both whistleblowers and multi-party compliance attestations, though it is not a verification mechanism in its own right. Training and debriefings are commonplace in fields that use sensitive information; for example, under the U.S. National Industrial Security Program, cleared contractor employees must complete insider-threat training.

\textbf{Recordkeeping and disclosure on request} Require high-risk organizations to maintain an auditable record of important activities and submit select records to inspectors upon request. This is easier and likely more palatable to implement than embedded auditors, but offers weaker evidence about compliance. Required documentation might include meeting notes, code changes, compute workloads, and other important records. Verifiable outputs from the organization (such as product releases) may serve as an additional point of comparison for validating records. Alternatively, a regime might require a cryptographically-signed ledger of high-risk organizations’ software code, recorded for later audit \citep{philippe_cryptographic_2019}. In a direct precedent for mandatory recordkeeping and disclosure, the CWC requires state parties to declare relevant activities to the OPCW and provide access to facility records during inspections.

\textbf{Evaluating AI models to detect unexplained capability gains} Perform third-party evaluations of AI models. The agreement from \citet{scher_international_2025} would constrain compute and research in order to prevent AI capabilities from improving. Therefore, capability increases would indicate that progress is being made in at least one of these areas. This mechanism is useful if there is low visibility into AI development for verification, for instance due to low political will. Verification authorities would need access to any model they intend to evaluate, meaning this mechanism cannot function when development is entirely secret, though it could function if authorities are granted early access to models before deployment. A further challenge is that the field of AI presently lacks a strong basis for predicting how inputs to AI development will affect downstream model capabilities \citep{ganguli_predictability_2022}, and also lacks evaluation techniques that are robust to increasingly test-aware AI models \citep{anthropic_system_2026,msl_preparedness__red_teaming__alignment_team_muse_2026}. Thus, measured model performance cannot provide definitive evidence, only indicate a need for further investigation.

\textbf{Controlled work environments to prevent evasion} Augment employee-monitoring software with methods that hinder evasion. For instance, require that high-risk researchers and high-risk organizations work in Sensitive Compartmented Information Facilities or other secured environments without access to personal devices, and require work devices to run monitoring software. Draw from existing companies’ efforts to ensure their employees are not secretly working a second job: monitoring network activity, badge/key entry logs, and more. This mechanism is useful if states are concerned that monitoring alone will not suffice to prevent high-risk researchers from covertly performing restricted research.

\subsection{Chip-use verification for declared chips}

The agreement proposed by \citet{scher_international_2025} involves locating and monitoring a large fraction of the world's AI chips, aiming for all non-trivial clusters of chips to be located. We refer readers to previous work for mechanisms that could help locate AI chips \citep{scher_international_2025,baker_verifying_2025,wasil_verification_2024,scher_mechanisms_2024,brass_location_2024,fist_preventing_2023,petrie_near-term_2024}. The agreement then applies chip-use verification protocols to ensure that chips are only doing permitted activities such as inference on existing models, but not prohibited activities such as large-scale training. A wide array of previous work aims to identify or build these chip-use verification methods \citep{baker_verifying_2025,scher_mechanisms_2024,harack_verification_2025,shavit_what_2023,heim_governing_2024}. In this section, we assume the vast majority of AI chips have been located and are being monitored, and we ask what specific claims about the chips would be relevant to ensuring restricted research is not taking place. The effectiveness of these mechanisms depends on how much coverage the chip monitoring system has over the total stock of relevant chips---a question we leave to other work. Similar to the mechanisms for declared high-risk organizations, these mechanisms involve a known and legitimate prover (i.e., the chip owner) cooperating with verification authorities.

\textbf{Review AI training for novel methods of relevance} Require that training on monitored compute clusters (this could be all training or all training above some compute scale, such as the \textit{monitored threshold} in \citet{scher_international_2025}'s agreement) be submitted to inspectors, including all code used, and review this training to ensure it does not use novel AI methods of relevance. Under an agreement, \textit{large} compute clusters are very likely to be known and monitored, whereas covert violators might have \textit{small} secret data centers that are unmonitored. These violators would prefer to use monitored clusters to help them with their AI development because these are major sources of compute. For example, they might conduct restricted research using their secret clusters, and then do larger-scale AI training using the larger (monitored) clusters because training is compute-intensive. Because these larger clusters are monitored, it would be possible to detect them doing so, specifically by reviewing the AI training on monitored clusters to determine if it uses novel methods of relevance (the results of restricted research). This check would ensure that all the AI algorithms used were either previously known or irrelevant to frontier AI, and the code and algorithms could be reviewed in an automated fashion to preserve privacy and security while ensuring compliance. Given that the mechanism can only be applied to monitored chips, it cannot preclude algorithmic progress being conducted in secret data centers and applied to training in secret data centers. Even with that limitation, this mechanism adds value by preventing new algorithms from being applied to monitored chips, a category which, under most plausible restriction regimes, includes all of the largest data centers.

\textbf{Monitoring AI inference content for restricted research} Apply classifiers to the content of all inference workloads (user prompts, tool use, model outputs) to detect restricted research. If chips are running inference on an AI model that is capable of restricted research, it is necessary to ensure that the model is not being used for this purpose. One difficulty is that AI models or users might obfuscate the workload, tricking the classifier into thinking the workload is permitted when it is not, similar to jailbreaking strategies that bypass external classifiers \citep{davies_boundary_2026}. Another difficulty is that this approach can only be applied to monitored chips; because inference requires relatively few chips \citep{deepseek_ai_day_2025}, this mechanism would only obtain high confidence if virtually all of the relevant chip clusters were monitored and had this classification applied. While many AI companies already deploy classifiers on their own inference workloads to detect prohibited uses \citep{sharma_constitutional_2025,openai_preparing_2025}, there would be a small difficulty in adding AI R\&D to the set of prohibited uses and a large difficulty in proving comprehensive monitoring to external inspectors.

\textbf{Review AI training data for prohibited data improvements} Log and review AI training data to verify the data complies with research restrictions. Improvements to the data used to train AIs can contribute to dangerous capabilities, and thus might be restricted by an agreement, but such improvements could be difficult to detect. \citet{choi_tools_2023} propose methods of verifying that certain data was used to train a particular model, but further research is needed to make these methods implementation-ready. Given such methods, however, inspectors could then attempt to confirm whether the verified training data complies with restrictions. For example, if there are rules limiting the use of synthetic data, inspectors could apply AI-text-detection methods to the training data. An agreement may find it challenging to define and classify data improvements; potential solutions may include restricting the amount of AI-generated data, restricting which models can be used for synthetic data generation, or restricting what domains the data covers. Current AI development involves classifying and filtering training data, for instance to remove CSAM, but this filtering is typically not verified by an external party.

\textbf{Broad chip-use monitoring for restricted experiments} Classify the activities of AI chips to detect restricted research. Many algorithmic advances require experiments to test them. This experimentation would be prohibited by an agreement, and chip-use verification could be applied to ensure chips are not running such experiments. For instance, automated classifiers could analyze characteristics of the workloads being run---power draw, memory bandwidth utilization, user organization---to determine if these are likely part of a restricted research project \citep{tang_mit_2022,rahman_detecting_2026}. If restrictions prohibit new AI training, classification could be aimed at differentiating training from inference. However, most realistic definitions of restricted research would not be so simple to enforce via classification of chip activities---ease of classification depends on how restrictions are defined. Classification can only cover monitored chips, and due to the relatively low number of chips required for research, this monitoring should be widely applied to make this approach work well. In a structurally similar practice, cloud providers offer tools for detecting unauthorized cryptocurrency mining \citep{google_cryptomining_2026}.

\textbf{Restricting chips to only run approved models} Ensure that AI chips do not run models capable of restricted research. This could involve an allowlist (approve only certain models), a blocklist (prohibit only certain models), or some other method of differentiating models (e.g., allow models below a certain size). Because AI capabilities are difficult to verify, it is hard to rule out that a model might conduct restricted research, and this state of affairs suggests an allowlist. Identifying models for an allowlist might involve data filtering or ``unlearning'' to limit AI skill at restricted research, training models to refuse it, or evaluating models to confirm they cannot perform it. At present, all of these methods face significant challenges in robustness, but they may nonetheless add value. Further challenges emerge in requiring chips to attest to the identity of models being run, given the proprietary nature of current AI software and hardware stacks, and like many alternatives, this mechanism can only be applied to monitored chips. The closest precedent is Digital Rights Management (DRM), which uses cryptographic checks to prevent unapproved uses of digital media or software, but DRM has a poor track record \citep{biddle_darknet_2003,patat_exploring_2022}.

\subsection{Non-proliferation}

Some mechanisms broadly aim at preventing dangerous resources or knowledge from reaching bad actors. Prevention is typically not labeled as a form of verification, but we discuss these non-proliferation options because they would correctly cause a verifying party to have greater assurance that rules are being followed. In the context of AI governance, non-proliferation is typically applied to AI chips, for instance via export controls. Chip-level non-proliferation would be important for verifying an international agreement, but here we defer to prior work on the topic \citep{fist_preventing_2023,allen_choking_2022,scher_international_2025}, and instead focus on mechanisms more uniquely relevant to verifying research restrictions.

\textbf{Preventing the development of AI-R\&D-capable AIs} Prevent, as a first line of defense, the development of AIs advanced enough to automate restricted research. Prior to an international agreement, domestic regulation could delay or partially limit such development. With an agreement in place, AI development can be restricted more effectively. An agreement might require that AI training have no more than 5\% of its training data be coding tokens and that the final model score below some threshold on AI R\&D benchmarks. Data use could be verified by a combination of training-data identification and classification as code or non-code. AI R\&D potential could be verified by evaluations, which are likely to include fine-tuning and other elicitation strategies; note, however, that upper bounding an AI's capabilities is not currently feasible \citep{barnett_what_2024}. This mechanism complements restrictions on the models that AI chips may run, as it seeks to prevent dangerous models from being developed at all.

\textbf{Preventing the proliferation of AI-R\&D-capable AIs} Prevent, as a second line of defense, the spread of AI systems that could automate restricted research. Under most plausible regimes, automating restricted research would require both unmonitored chips and the weights of a sufficiently capable AI\@. Access to the latter might be limited by heightened security in AI companies, regulations and international norms around releasing model weights, and explicit corporate non-proliferation policies. Note, however, that corporate security is unlikely to stand up to attacks by well-resourced state actors. This mechanism complements restrictions on the models that AI chips may run, as it seeks to prevent the proliferation of dangerous models. The Wassenaar Arrangement's 2013 addition of ``intrusion software'' to its dual-use control list is the closest precedent for multilateral export controls on a category of software \citep{wassenaar_arrangement_list_2013}.

\textbf{Non-proliferation of AI algorithms and data} Secure the algorithms and data available to AI research organizations today. To the extent that groups capable of frontier AI research do not publish or leak their methods, knowledge proliferation will be limited to what can be gleaned from the open-source community, perhaps several months behind the frontier. Non-proliferation could apply to both algorithmic secrets and to data that advances AI capabilities (e.g., high-quality RL environments). Even if state actors have extracted some secrets, these secrets might lag the frontier, and these state actors could implement their own security against further proliferation. This mechanism would initially delay restricted research projects, but it would not provide ongoing assurance that such projects have not advanced. One difficulty is the fact that algorithmic secrets are often small pieces of knowledge that are hard to secure. The Atomic Energy Act's ``born secret'' doctrine provides precedent for non-proliferation of research ideas, as it renders information pertaining to the design of nuclear weapons classified by default \citep{united_states_congress_atomic_1954}.

\section{Discussion and Conclusion}

We have analyzed both key considerations affecting the verifiability of research restrictions and specific high-level mechanisms that could be used to implement those restrictions. This analysis leads to three main takeaways.

First, the most difficult challenge is the risk of secret data centers that are not known to verification authorities. These data centers could be used for experimentation (testing research ideas as part of the illicit research process) and for unmonitored inference (AIs thinking through research ideas and writing code for experiments). This consideration is especially important if AI models become sufficiently capable at AI R\&D tasks before an agreement. Before that point, both human experts and compute are necessary components for most AI R\&D, but after that point, only compute is needed.

Second, a few mechanisms stand out as especially useful. Whistleblowers and intelligence gathering are well established, and would likely be the most effective methods for detecting covert projects. Meanwhile, the best methods for monitoring declared organizations would be automated methods, such as automated code reviews and inference monitoring. These methods would aim to identify illicit experiments or AIs thinking about and writing code for restricted research. They could be designed to uphold privacy and security by default, only escalating to human review when needed. Also especially valuable is reviewing AI training for novel methods of relevance, which may prevent covert projects from using monitored chips to scale any restricted research they conduct.

Third, available mechanisms vary widely in ease of implementation. Whistleblowers, interviews, traditional intelligence gathering, and multi-party compliance attestations would be straightforward to implement, while some mechanisms first require significant R\&D, especially chip-use verification.

One major risk related to the mechanisms in this report is that, if implemented without adequate safeguards, they could be misused by bad actors for purposes such as political repression, suppression of legitimate research, surveillance overreach, and security leakage. Some mechanisms, such as whistleblowers, are minimally invasive and pose little risk of co-option, while others would require extremely careful implementation to avoid misuse. Additionally, well-meaning actors could implement restrictions poorly, causing collateral damage to unrestricted research or other harms. We hope future work builds implementation-ready versions of some of the mechanisms in this report while carefully considering security, privacy, and efficacy.

Overall, we remain uncertain about how easy it will be for countries to verify restrictions on frontier AI research in an international agreement; there are points favoring both optimism and pessimism. Insofar as it is necessary to have research restrictions and verify compliance with them, this work advances the state of thinking on the topic by highlighting important considerations and potential mechanisms---some of which require urgent R\&D if they are to be ready soon.

\section*{Impact statement}

This work aims to advance the field of technical AI governance by exploring verification methods. This topic is important to the viability of international agreements, and providing more detailed proposals for verification can make conversations about such agreements more concrete. There are risks from this work, including that verification mechanisms could be misappropriated toward harmful ends, and that research on verification could serve as a guide for evaders. We believe our counterfactual impact on both of these is sufficiently small that it is overwhelmed by the benefit of creating more detailed AI governance proposals.

\section*{LLM usage statement}

LLMs were primarily used to convert this paper into \LaTeX{} and to review drafts for typos. They were also used occasionally in brainstorming, locating citations, or to assist in rewriting particular sentences. The vast majority of the ideas in this paper and final text were human-contributed.

\bibliography{Research_Restriction_Verification}

@misc{petrie_near-term_2024,
	title = {Near-{Term} {Enforcement} of {AI} {Chip} {Export} {Controls} {Using} {A} {Firmware}-{Based} {Design} for {Offline} {Licensing}},
	url = {http://arxiv.org/abs/2404.18308},
	doi = {10.48550/arXiv.2404.18308},
	urldate = {2024-10-28},
	publisher = {arXiv},
	author = {Petrie, James},
	month = may,
	year = {2024},
	note = {arXiv:2404.18308},
}

@article{philippe_cryptographic_2019,
	title = {A {Cryptographic} {Escrow} for {Treaty} {Declarations} and {Step}-by-{Step} {Verification}},
	volume = {27},
	issn = {0892-9882, 1547-7800},
	url = {https://www.tandfonline.com/doi/full/10.1080/08929882.2019.1573483},
	doi = {10.1080/08929882.2019.1573483},
	language = {en},
	number = {1},
	urldate = {2024-10-28},
	journal = {Science \& Global Security},
	author = {Philippe, Sébastien and Glaser, Alexander and Felten, Edward W.},
	month = jan,
	year = {2019},
	pages = {3--14},
}

@techreport{fist_preventing_2023,
	title = {Preventing {AI} {Chip} {Smuggling} to {China}},
	url = {https://www.cnas.org/publications/reports/preventing-ai-chip-smuggling-to-china},
	language = {en},
	urldate = {2024-10-29},
	institution = {Center for New American Security},
	author = {Fist, Tim and Grunewald, Erich},
	month = oct,
	year = {2023},
}

@misc{heim_governing_2024,
	title = {Governing {Through} the {Cloud}: {The} {Intermediary} {Role} of {Compute} {Providers} in {AI} {Regulation}},
	shorttitle = {Governing {Through} the {Cloud}},
	url = {http://arxiv.org/abs/2403.08501},
	urldate = {2024-10-31},
	publisher = {arXiv},
	author = {Heim, Lennart and Fist, Tim and Egan, Janet and Huang, Sihao and Zekany, Stephen and Trager, Robert and Osborne, Michael A. and Zilberman, Noa},
	month = mar,
	year = {2024},
	note = {arXiv:2403.08501},
}

@misc{choi_tools_2023,
	title = {Tools for {Verifying} {Neural} {Models}' {Training} {Data}},
	url = {http://arxiv.org/abs/2307.00682},
	urldate = {2024-11-01},
	publisher = {arXiv},
	author = {Choi, Dami and Shavit, Yonadav and Duvenaud, David},
	month = jul,
	year = {2023},
	note = {arXiv:2307.00682},
}

@article{nagin_deterrence_2013,
	title = {Deterrence: {A} {Review} of the {Evidence} by a {Criminologist} for {Economists}},
	volume = {5},
	issn = {1941-1383, 1941-1391},
	shorttitle = {Deterrence},
	url = {https://www.annualreviews.org/content/journals/10.1146/annurev-economics-072412-131310},
	doi = {10.1146/annurev-economics-072412-131310},
	language = {en},
	number = {Volume 5, 2013},
	urldate = {2026-04-25},
	journal = {Annual Review of Economics},
	publisher = {Annual Reviews},
	author = {Nagin, Daniel S.},
	month = aug,
	year = {2013},
	pages = {83--105},
}

@misc{mark_constitutional_2025,
	title = {Constitutional {Law} and {AI} {Governance}: {Constraints} on {Model} {Licensing} and {Research} {Classification}},
	shorttitle = {Constitutional {Law} and {AI} {Governance}},
	url = {https://arxiv.org/abs/2509.05361v2},
	language = {en},
	urldate = {2026-04-25},
	journal = {arXiv.org},
	author = {Mark, Alex and Scher, Aaron},
	month = sep,
	year = {2025},
}

@misc{dean_jeff_2025,
	title = {Jeff {Dean} \& {Noam} {Shazeer} — 25 years at {Google}: from {PageRank} to {AGI}},
	url = {https://www.dwarkesh.com/p/jeff-dean-and-noam-shazeer},
	urldate = {2026-04-25},
	author = {Dean, Jeff and Shazeer, Noam},
	month = feb,
	year = {2025},
}

@misc{whitfill_will_2025,
	title = {Will {Compute} {Bottlenecks} {Prevent} an {Intelligence} {Explosion}?},
	url = {https://arxiv.org/abs/2507.23181v2},
	language = {en},
	urldate = {2026-04-25},
	journal = {arXiv.org},
	author = {Whitfill, Parker and Wu, Cheryl},
	month = jul,
	year = {2025},
}

@misc{gundlach_origin_2025,
	title = {On the {Origin} of {Algorithmic} {Progress} in {AI}},
	url = {https://arxiv.org/abs/2511.21622v1},
	language = {en},
	urldate = {2026-04-25},
	journal = {arXiv.org},
	author = {Gundlach, Hans and Fogelson, Alex and Lynch, Jayson and Trisovic, Ana and Rosenfeld, Jonathan and Sandhu, Anmol and Thompson, Neil},
	month = nov,
	year = {2025},
}

@misc{buchanan_ai_2020,
	title = {The {AI} {Triad} and {What} {It} {Means} for {National} {Security} {Strategy}},
	url = {https://cset.georgetown.edu/publication/the-ai-triad-and-what-it-means-for-national-security-strategy/},
	language = {en-US},
	urldate = {2026-04-25},
	journal = {Center for Security and Emerging Technology},
	author = {Buchanan, Ben},
	month = aug,
	year = {2020},
}

@misc{barnett_compute_2025,
	title = {Compute {Requirements} for {Algorithmic} {Innovation} in {Frontier} {AI} {Models}},
	url = {https://arxiv.org/abs/2507.10618v1},
	language = {en},
	urldate = {2026-04-25},
	journal = {arXiv.org},
	author = {Barnett, Peter},
	month = jul,
	year = {2025},
}

@misc{martin_analysis_2024,
	title = {Analysis of {Global} {AI} {Governance} {Strategies} {\textbar} {Convergence} {Analysis}},
	url = {https://www.convergenceanalysis.org/research/analysis-of-global-ai-governance-strategies},
	language = {en},
	urldate = {2026-04-25},
	author = {Martin, Sammy and Bullock, Justin and Katzke, Corin},
	month = dec,
	year = {2024},
}

@misc{ramiah_toward_2025,
	title = {Toward a {Global} {Regime} for {Compute} {Governance}: {Building} the {Pause} {Button}},
	shorttitle = {Toward a {Global} {Regime} for {Compute} {Governance}},
	url = {https://arxiv.org/abs/2506.20530v1},
	language = {en},
	urldate = {2026-04-25},
	journal = {arXiv.org},
	author = {Ramiah, Ananthi Al and Koopmanschap, Raymond and Thorsteinson, Josh and Khan, Sadruddin and Zhou, Jim and Noh, Shafira and Meindertsma, Joep and Shafiq, Farhan},
	month = jun,
	year = {2025},
}

@misc{center_for_ai_safety_statement_2023,
	title = {Statement on {AI} {Risk} {\textbar} {CAIS}},
	url = {https://aistatement.com},
	language = {en},
	urldate = {2026-04-25},
	journal = {Center for AI Safety},
	author = {{Center for AI Safety}},
	year = {2023},
}

@book{yudkowsky_if_2025,
	address = {New York},
	title = {If {Anyone} {Builds} {It}, {Everyone} {Dies}: {Why} {Superhuman} {AI} {Would} {Kill} {Us} {All}},
	isbn = {978-0-316-59564-3},
	shorttitle = {If {Anyone} {Builds} {It}, {Everyone} {Dies}},
	language = {English},
	publisher = {Little, Brown and Company},
	author = {Yudkowsky, Eliezer and Soares, Nate},
	year = {2025},
}

@misc{scher_international_2025,
	title = {An {International} {Agreement} to {Prevent} the {Premature} {Creation} of {Artificial} {Superintelligence}},
	url = {https://arxiv.org/abs/2511.10783v2},
	language = {en},
	urldate = {2026-04-25},
	author = {Scher, Aaron and Abecassis, David and Barnett, Peter and Abeyta, Brian},
	month = nov,
	year = {2025},
}

@misc{barnett_ai_2025,
	title = {{AI} {Governance} to {Avoid} {Extinction}: {The} {Strategic} {Landscape} and {Actionable} {Research} {Questions}},
	shorttitle = {{AI} {Governance} to {Avoid} {Extinction}},
	url = {https://arxiv.org/abs/2505.04592v1},
	language = {en},
	urldate = {2026-04-25},
	journal = {arXiv.org},
	author = {Barnett, Peter and Scher, Aaron},
	month = may,
	year = {2025},
}

@techreport{committee_of_privy_counsellors_review_2004,
	address = {London},
	title = {Review of {Intelligence} on {Weapons} of {Mass} {Destruction}},
	institution = {The Stationery Office},
	author = {{Committee of Privy Counsellors}},
	year = {2004},
}

@book{national_research_council_polygraph_2003,
	address = {Washington, D.C},
	title = {The {Polygraph} and {Lie} {Detection}},
	isbn = {978-0-309-08436-9},
	language = {English},
	publisher = {National Academies Press},
	author = {{National Research Council}},
	year = {2003},
}

@misc{schlesinger_memorandum_1973,
	title = {Memorandum {From} {Director} of {Central} {Intelligence} {Schlesinger} to {All} {Central} {Intelligence} {Agency} {Employees}},
	url = {https://www.cia.gov/readingroom/docs/memo%20for%20all%20cia%20employee%5B15132481%5D.pdf},
	publisher = {Central Intelligence Agency},
	author = {Schlesinger, James},
	month = may,
	year = {1973},
}

@misc{us_nuclear_regulatory_commission_resident_2025,
	title = {Resident {Inspector} {Program} {Overview}},
	url = {https://www.nrc.gov/reactors/operating/oversight/rop-description/resident-insp-program},
	author = {{U.S. Nuclear Regulatory Commission}},
	year = {2025},
}

@misc{united_states_congress_sarbanes-oxley_2002,
	chapter = {302},
	title = {Sarbanes-{Oxley} {Act} of 2002},
	author = {{United States Congress}},
	month = jul,
	year = {2002},
}

@misc{baker_verifying_2025,
	title = {Verifying {International} {Agreements} on {AI}: {Six} {Layers} of {Verification} for {Rules} on {Large}-{Scale} {AI} {Development} and {Deployment}},
	shorttitle = {Verifying {International} {Agreements} on {AI}},
	url = {https://arxiv.org/abs/2507.15916v2},
	language = {en},
	urldate = {2026-04-25},
	journal = {arXiv.org},
	author = {Baker, Mauricio and Kulp, Gabriel and Marks, Oliver and Brundage, Miles and Heim, Lennart},
	month = jul,
	year = {2025},
}

@misc{wasil_verification_2024,
	title = {Verification methods for international {AI} agreements},
	url = {https://arxiv.org/abs/2408.16074v2},
	language = {en},
	urldate = {2026-04-25},
	journal = {arXiv.org},
	author = {Wasil, Akash R. and Reed, Tom and Miller, Jack William and Barnett, Peter},
	month = aug,
	year = {2024},
}

@misc{scher_mechanisms_2024,
	title = {Mechanisms to {Verify} {International} {Agreements} {About} {AI} {Development}},
	url = {https://arxiv.org/abs/2506.15867},
	urldate = {2026-04-25},
	author = {Scher, Aaron and Thiergart, Lisa},
	month = nov,
	year = {2024},
}

@misc{brass_location_2024,
	title = {Location {Verification} for {AI} {Chips}},
	url = {https://www.iaps.ai/research/location-verification-for-ai-chips},
	language = {en-US},
	urldate = {2026-04-25},
	journal = {Institute for AI Policy and Strategy},
	author = {Brass, Asher and Aarne, Onni},
	month = apr,
	year = {2024},
}

@misc{shavit_what_2023,
	title = {What does it take to catch a {Chinchilla}? {Verifying} {Rules} on {Large}-{Scale} {Neural} {Network} {Training} via {Compute} {Monitoring}},
	shorttitle = {What does it take to catch a {Chinchilla}?},
	url = {https://arxiv.org/abs/2303.11341v2},
	language = {en},
	urldate = {2026-04-25},
	journal = {arXiv.org},
	author = {Shavit, Yonadav},
	month = mar,
	year = {2023},
}

@misc{ecoffet_event_2026,
	title = {Event {Replay}: {Sam} {Altman} on {Building} the {Future} of {AI} - {Video}},
	shorttitle = {Event {Replay}},
	url = {https://forum.openai.com/public/videos/event-replay-sam-altman-on-building-the-future-of-ai-2026-04-06},
	language = {en},
	urldate = {2026-04-25},
	journal = {OpenAI Forum},
	author = {Ecoffet, Adrien and Altman, Sam and Achiam, Joshua and Nicholson, Chris},
	month = apr,
	year = {2026},
}

@misc{anthropic_system_2026,
	title = {System {Card}: {Claude} {Mythos} {Preview}},
	url = {https://www-cdn.anthropic.com/08ab9158070959f88f296514c21b7facce6f52bc.pdf},
	author = {{Anthropic}},
	month = apr,
	year = {2026},
}

@article{davidson_will_2025,
	title = {Will compute bottlenecks prevent a software intelligence explosion? — {AI} {Alignment} {Forum}},
	shorttitle = {Will compute bottlenecks prevent a software intelligence explosion?},
	url = {https://www.alignmentforum.org/posts/XDF6ovePBJf6hsxGj/will-compute-bottlenecks-prevent-a-software-intelligence-1},
	urldate = {2026-04-25},
	author = {Davidson, Tom},
	month = apr,
	year = {2025},
}

@misc{barnett_what_2024,
	title = {What {AI} evaluations for preventing catastrophic risks can and cannot do},
	url = {http://arxiv.org/abs/2412.08653},
	doi = {10.48550/arXiv.2412.08653},
	urldate = {2026-04-25},
	publisher = {arXiv},
	author = {Barnett, Peter and Thiergart, Lisa},
	month = nov,
	year = {2024},
	note = {arXiv:2412.08653 [cs]},
}

@misc{davies_boundary_2026,
	title = {Boundary {Point} {Jailbreaking} of {Black}-{Box} {LLMs}},
	url = {http://arxiv.org/abs/2602.15001},
	doi = {10.48550/arXiv.2602.15001},
	urldate = {2026-04-25},
	publisher = {arXiv},
	author = {Davies, Xander and Giglemiani, Giorgi and Lau, Edmund and Winsor, Eric and Irving, Geoffrey and Gal, Yarin},
	month = feb,
	year = {2026},
	note = {arXiv:2602.15001 [cs]},
}

@misc{sharma_constitutional_2025,
	title = {Constitutional {Classifiers}: {Defending} against {Universal} {Jailbreaks} across {Thousands} of {Hours} of {Red} {Teaming}},
	shorttitle = {Constitutional {Classifiers}},
	url = {http://arxiv.org/abs/2501.18837},
	doi = {10.48550/arXiv.2501.18837},
	urldate = {2026-04-25},
	publisher = {arXiv},
	author = {Sharma, Mrinank and Tong, Meg and Mu, Jesse and Wei, Jerry and Kruthoff, Jorrit and Goodfriend, Scott and Ong, Euan and Peng, Alwin and Agarwal, Raj and Anil, Cem and Askell, Amanda and Bailey, Nathan and Benton, Joe and Bluemke, Emma and Bowman, Samuel R. and Christiansen, Eric and Cunningham, Hoagy and Dau, Andy and Gopal, Anjali and Gilson, Rob and Graham, Logan and Howard, Logan and Kalra, Nimit and Lee, Taesung and Lin, Kevin and Lofgren, Peter and Mosconi, Francesco and O'Hara, Clare and Olsson, Catherine and Petrini, Linda and Rajani, Samir and Saxena, Nikhil and Silverstein, Alex and Singh, Tanya and Sumers, Theodore and Tang, Leonard and Troy, Kevin K. and Weisser, Constantin and Zhong, Ruiqi and Zhou, Giulio and Leike, Jan and Kaplan, Jared and Perez, Ethan},
	month = jan,
	year = {2025},
	note = {arXiv:2501.18837 [cs]},
}

@misc{openai_preparing_2025,
	title = {Preparing for future {AI} capabilities in biology},
	url = {https://openai.com/index/preparing-for-future-ai-capabilities-in-biology/},
	language = {en-US},
	urldate = {2026-04-25},
	journal = {OpenAI},
	author = {{OpenAI}},
	year = {2025},
}

@inproceedings{biddle_darknet_2003,
	title = {The {Darknet} and the {Future} of {Content} {Distribution}},
	volume = {6},
	url = {https://www.semanticscholar.org/paper/The-Darknet-and-the-Future-of-Content-Distribution-Biddle-England/d09ea4f3bb7a0b8ffc596eaa92af8ae367485358},
	urldate = {2026-04-25},
	booktitle = {{ACM} {Workshop} on digital rights management},
	author = {Biddle, Peter and England, Paul and Peinado, Marcus and Willman, Bryan},
	month = oct,
	year = {2003},
	pages = {54},
}

@inproceedings{patat_exploring_2022,
	title = {Exploring {Widevine} for {Fun} and {Profit}},
	issn = {2770-8411},
	url = {https://ieeexplore.ieee.org/abstract/document/9833867},
	doi = {10.1109/SPW54247.2022.9833867},
	urldate = {2026-04-25},
	booktitle = {2022 {IEEE} {Security} and {Privacy} {Workshops} ({SPW})},
	author = {Patat, Gwendal and Sabt, Mohamed and Fouque, Pierre-Alain},
	month = may,
	year = {2022},
	note = {ISSN: 2770-8411},
	pages = {277--288},
}

@techreport{allen_choking_2022,
	title = {Choking off {China}’s {Access} to the {Future} of {AI}},
	url = {https://www.csis.org/analysis/choking-chinas-access-future-ai},
	language = {en},
	urldate = {2026-04-25},
	author = {Allen, Gregory C.},
	month = oct,
	year = {2022},
}

@article{harack_verification_2025,
	title = {Verification for international {AI} governance},
	journal = {AI Governance Initiative, Oxford Martin School, University of Oxford},
	author = {Harack, Ben and Trager, Robert F and Reuel, Anka and Manheim, David and Brundage, Miles and Aarne, Onni and Scher, Aaron and Pan, Yanliang and Xiao, Jenny and Loke, Kristy},
	year = {2025},
}

@misc{ho_algorithmic_2024,
	title = {Algorithmic progress in language models},
	url = {http://arxiv.org/abs/2403.05812},
	doi = {10.48550/arXiv.2403.05812},
	urldate = {2026-04-25},
	publisher = {arXiv},
	author = {Ho, Anson and Besiroglu, Tamay and Erdil, Ege and Owen, David and Rahman, Robi and Guo, Zifan Carl and Atkinson, David and Thompson, Neil and Sevilla, Jaime},
	month = mar,
	year = {2024},
	note = {arXiv:2403.05812 [cs]},
}

@misc{hooker_limitations_2024,
	title = {On the {Limitations} of {Compute} {Thresholds} as a {Governance} {Strategy}},
	url = {http://arxiv.org/abs/2407.05694},
	doi = {10.48550/arXiv.2407.05694},
	urldate = {2026-04-25},
	publisher = {arXiv},
	author = {Hooker, Sara},
	month = jul,
	year = {2024},
	note = {arXiv:2407.05694 [cs]},
}

@misc{heim_training_2024,
	title = {Training {Compute} {Thresholds}: {Features} and {Functions} in {AI} {Regulation}},
	shorttitle = {Training {Compute} {Thresholds}},
	url = {http://arxiv.org/abs/2405.10799},
	doi = {10.48550/arXiv.2405.10799},
	urldate = {2026-04-25},
	publisher = {arXiv},
	author = {Heim, Lennart and Koessler, Leonie},
	month = aug,
	year = {2024},
	note = {arXiv:2405.10799 [cs]},
}

@techreport{msl_preparedness__red_teaming__alignment_team_muse_2026,
	title = {Muse {Spark} {Safety} \& {Preparedness} {Report}},
	url = {https://ai.meta.com/static-resource/muse-spark-safety-and-preparedness-report/},
	author = {{MSL Preparedness \& Red Teaming \& Alignment Team} and {MSL AI Security Team}},
	month = apr,
	year = {2026},
}

@inproceedings{ganguli_predictability_2022,
	title = {Predictability and surprise in large generative models},
	author = {Ganguli, Deep and Hernandez, Danny and Lovitt, Liane and Askell, Amanda and Bai, Yuntao and Chen, Anna and Conerly, Tom and Dassarma, Nova and Drain, Dawn and Elhage, Nelson},
	year = {2022},
	pages = {1747--1764},
}

@inproceedings{tang_mit_2022,
	title = {The {MIT} {Supercloud} {Workload} {Classification} {Challenge}},
	isbn = {1-6654-9747-5},
	publisher = {IEEE},
	author = {Tang, Benny J and Chen, Qiqi and Weiss, Matthew L and Frey, Nathan C and McDonald, Joseph and Bestor, David and Yee, Charles and Arcand, William and Bergeron, William and Byun, Chansup},
	year = {2022},
	pages = {708--714},
}

@misc{google_cryptomining_2026,
	title = {Cryptomining detection best practices {\textbar} {Security} {Command} {Center}},
	url = {https://docs.cloud.google.com/security-command-center/docs/cryptomining-detection-best-practices},
	language = {en},
	urldate = {2026-04-25},
	journal = {Google Cloud Documentation},
	author = {{Google}},
	month = apr,
	year = {2026},
}

@misc{future_of_life_institute_statement_2025,
	title = {Statement on {Superintelligence}},
	url = {https://superintelligence-statement.org},
	language = {en},
	urldate = {2026-06-24},
	journal = {Statement on Superintelligence},
	author = {{Future of Life Institute}},
	month = sep,
	year = {2025},
}

@misc{reedi_ml-intern_2026,
	title = {ml-intern: an agent that autonomously researches, writes, and ships good quality {ML} related code using the {Hugging} {Face} ecosystem},
	url = {https://github.com/huggingface/ml-intern},
	author = {Reedi, Aksel Joonas and Bonamy, Henri and Di Cosmo, Yoan and von Werra, Leandro and Tunstall, Lewis},
	year = {2026},
}

@misc{deepseek_ai_day_2025,
	title = {Day 6: {One} {More} {Thing}, {DeepSeek}-{V3}/{R1} {Inference} {System} {Overview}},
	url = {https://github.com/deepseek-ai/open-infra-index/blob/main/202502OpenSourceWeek/day_6_one_more_thing_deepseekV3R1_inference_system_overview.md},
	language = {en},
	urldate = {2026-06-25},
	journal = {GitHub},
	author = {{DeepSeek AI}},
	month = feb,
	year = {2025},
}

@misc{wassenaar_arrangement_list_2013,
	title = {List of {Dual}-{Use} {Goods} and {Technologies} and {Munitions} {List}},
	url = {https://www.wassenaar.org/app/uploads/2019/consolidated/WA-LIST%20%2813%29%201.pdf},
	author = {{Wassenaar Arrangement}},
	month = dec,
	year = {2013},
	note = {Published: WA-LIST (13) 1},
}

@misc{united_states_congress_atomic_1954,
	title = {Atomic {Energy} {Act} of 1954},
	author = {{United States Congress}},
	year = {1954},
	note = {Published: 42 U.S.C. \S{} 2014(y)},
}

@misc{favaro_when_2026,
	title = {When {AI} builds itself},
	url = {https://www.anthropic.com/institute/recursive-self-improvement},
	language = {en},
	urldate = {2026-06-25},
	author = {Favaro, Marina and Clark, Jack},
	month = jun,
	year = {2026},
}

@misc{altman_built_2026,
	title = {Built to benefit everyone: our plan},
	shorttitle = {Built to benefit everyone},
	url = {https://openai.com/index/built-to-benefit-everyone-our-plan/},
	language = {en-US},
	urldate = {2026-06-25},
	journal = {OpenAI},
	author = {Altman, Sam and Pachocki, Jakub},
	month = jun,
	year = {2026},
}

@misc{united_states_congress_invention_1952,
	title = {Invention {Secrecy} {Act} of 1951},
	url = {https://www.govinfo.gov/link/statute/66/3},
	author = {{United States Congress}},
	month = feb,
	year = {1952},
	note = {Published: Pub. L. No. 82-256, ch. 4, 66 Stat. 3},
}

@inproceedings{rahman_detecting_2026,
	title = {Detecting {Hidden} {ML} {Training} {With} {Zero}-{Overhead} {Telemetry}},
	url = {https://arxiv.org/abs/2606.19262},
	booktitle = {Technical {AI} {Governance} {Research} {Workshop} at the 43rd {International} {Conference} on {Machine} {Learning} ({ICML})},
	author = {Rahman, Robi and Tajdari, Sabiha},
	year = {2026},
}
\bibliographystyle{icml2026}

\end{document}